\documentclass[floats,aps,showpacs,twocolumn,superscriptaddress,pre]{revtex4-1}

\usepackage[dvips]{graphicx}
\usepackage{amsfonts}
\usepackage{amsmath,amssymb}
\usepackage{dsfont}
\usepackage{placeins}
\usepackage{afterpage}
\usepackage{flafter} 

\newcommand{\Nbb}{N_{\text{bb}}}

\newcommand{\Int}{\int\limits}
\newcommand{\ud}{\text{d}}

\newcommand{\psiregW}{\psi_{\text{reg},W}}

\newcommand{\ue}{\text{e}}

\newcommand{\psiregmn}{\psi_{\text{reg}}^{mn}}

\newcommand{\psiregmsns}{\psi_{\text{reg}}^{m'n'}}

\newcommand{\psich}{\psi_{\text{ch}}}

\newcommand{\PdE}{P_{\Delta E}}
\newcommand{\PchdE}{P_{\text{ch},\Delta E}}

\newcommand{\Hreg}{H_{\text{reg}}}

\newcommand{\Eregmn}{E_{\text{reg}}^{mn}}

\newcommand{\vchmn}{v_{\text{ch},mn}}
\newcommand{\vmn}{\bar{v}_{mn}}
\newcommand{\vmnmsns}{v_{m'n',mn}}

\newcommand{\opone}{\mathds{1}}

\newcommand{\cosb}{\mathcal{B}_c}
\newcommand{\stadb}{\mathcal{B}_s}

\begin{document}

\title{Coupling of bouncing-ball modes to the chaotic sea and their counting function}

\author{Steffen L\"ock}
\affiliation{Institut f\"ur Theoretische Physik, Technische Universit\"at
             Dresden, 01062 Dresden, Germany}

\author{Arnd B\"acker}
\affiliation{Institut f\"ur Theoretische Physik, Technische Universit\"at
             Dresden, 01062 Dresden, Germany}
\affiliation{Max-Planck-Institut f\"ur Physik komplexer Systeme, N\"othnitzer
Stra\ss{}e 38, 01187 Dresden, Germany}

\author{Roland Ketzmerick}       
\affiliation{Institut f\"ur Theoretische Physik, Technische Universit\"at
             Dresden, 01062 Dresden, Germany}
\affiliation{Max-Planck-Institut f\"ur Physik komplexer Systeme, N\"othnitzer
Stra\ss{}e 38, 01187 Dresden, Germany}

\date{\today}

\begin{abstract}
We study the coupling of bouncing-ball modes to chaotic modes in 
two-dimensional billiards with two parallel boundary segments.
Analytically, we predict the corresponding decay
rates using the fictitious integrable system approach. Agreement with 
numerically determined rates is found for the stadium and the cosine billiard. 
We use this result to predict the asymptotic behavior of the counting function 
$\Nbb(E)\sim E^\delta$. For the stadium billiard we 
find agreement with the previous result $\delta = 3/4$. 
For the cosine billiard we derive $\delta = 5/8$, which is 
confirmed numerically and is well below the 
previously predicted upper bound $\delta=9/10$.

\end{abstract}

\maketitle

\noindent


\section{Introduction}
\label{sec:intro}

Two-dimensional billiard systems have found abundant applications in 
contemporary physics such as for electromagnetic and acoustic resonators, 
microdisk lasers, atomic matter waves in optical billiards, and quantum 
dots \cite{Sto1999,NoeSto1997,FriKapCarDav2001,EllSchBer2001,
MarRimWesHopGos1992}.
The classical dynamics of these billiards is described by a point particle 
of mass $M$ moving with constant velocity inside a domain $\Omega$ with 
elastic reflections at its boundary $\partial\Omega$. Depending on the shape 
of the boundary the phase space can be regular, mixed regular-chaotic, or 
chaotic. Quantum mechanically, billiards are described by the time-independent 
Schr\"odinger equation (in units $\hbar=2M=1$)
\begin{eqnarray}
\label{eq:examples:billiards:quant:sgl}
 -\Delta \psi_l(x,y) = E_l\psi_l(x,y),\quad (x,y)\in\Omega
\end{eqnarray}
with Dirichlet boundary condition $\psi_l=0$ on $\partial\Omega$, eigenfunctions
$\psi_l$, eigenvalues $E_l$, Laplace operator $\Delta$, and $l\in\mathbb{N}$.

\begin{figure}[b]
  \begin{center}
     \includegraphics[width=85mm]{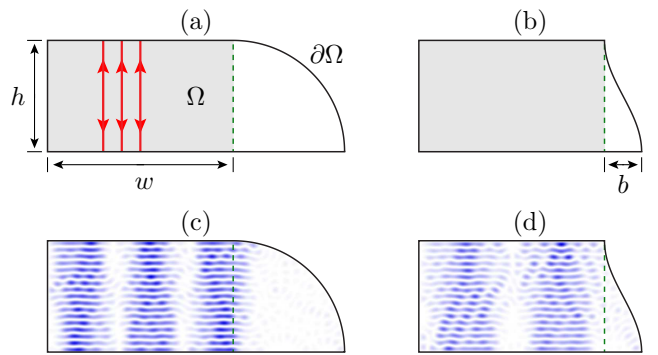}
     \caption{(Color online) Schematic pictures of the desymmetrized (a) 
              stadium and (b) cosine billiard. Each billiard has a rectangular 
              bouncing-ball region (gray shaded) of width $w$ and height $h$. 
              In the rectangular region so-called bouncing-ball 
              orbits exist perpendicular to the parallel parts of the boundary 
              (red vertical lines).
              In (c) and (d) bouncing-ball modes are shown for the stadium 
              and the cosine billiard, respectively.}
     \label{fig:billiards}
  \end{center}
\end{figure}

Many nonintegrable billiards of interest contain a rectangular region combined 
with other boundary segments, for example the stadium \cite{Bun1979}, 
the Sinai \cite{Sin1970}, and the cosine billiard \cite{Sti1996}; 
see Fig.~\ref{fig:billiards}. 
Classically, in these billiards a family of marginally stable periodic orbits 
exists, which bounce with perpendicular reflections between the two parallel 
parts of the boundary. These so-called bouncing-ball orbits are surrounded by 
chaotic motion in phase space. For simplicity we will only consider billiards
which show no visible regular regions.
Quantum mechanically, most of the eigenfunctions $\psi_l$ are chaotic; i.e., 
they extend over the whole chaotic phase space. 
In contrast, the so-called bouncing-ball modes $\psi_{mn}$ \cite{Hel1984,
BaiHosSteTay1985,OcoHel1988,McdKau1988,BogSch2004,BurZwo2005} concentrate on the 
marginally stable 
periodic orbits. They have a structure similar to the eigenstates of a 
rectangle; see Figs.~\ref{fig:billiards}(c) and \ref{fig:billiards}(d), 
and are characterized by the quantum 
numbers $m$ and $n$, where $m$ describes the quantization along the 
bouncing-ball orbits and $n$ perpendicular to them. 
In Ref.~\cite{Has2010} it was proven that semiclassically the modes with 
$n=1$ and increasing $m$ exist.  
For the counting function $\Nbb(E)$ of the
bouncing-ball modes up to energy $E$ this sequence of modes 
leads to $\Nbb(E)\sim\sqrt{E}$. Typically also modes with higher 
excitations $n>1$ perpendicular to the periodic orbits exist for large enough 
$m$. Which bouncing-ball modes $\psi_{mn}$ are realized
for a specific billiard depends on the couplings of the bouncing-ball modes to 
the chaotic modes. For fixed $m$ these couplings 
increase with $n$ such that for large $n$ the
bouncing-ball modes couple to many neighboring chaotic modes and thus disappear.
One expects that the number of bouncing-ball modes is asymptotically 
described by a power law $\Nbb(E) \propto E^\delta$ with exponent 
$1/2 \leq \delta <1$.
The exponent $\delta=1$ cannot be achieved in ergodic billiards, as quantum 
ergodicity \cite{Shn1974,Ver1985,Zel1987,GerLei1993,ZelZwo1996} 
requires that the fraction of exceptional 
eigenfunctions must vanish, i.e., $\Nbb(E)/N(E)\to 0$, where 
$N(E)\sim E$ is the total number of eigenstates.
 
The exponent $\delta$ depends on the shape of the billiard in the vicinity of 
the rectangular bouncing-ball region.
In Ref.~\cite{Tan1997} it was shown that for the stadium billiard the exponent 
$\delta=3/4$ arises, using an EBK-like quantization of the bouncing-ball modes.
With a different approach based on an adiabatic approximation 
\cite{BaiHosSteTay1985} of the bouncing-ball modes an upper bound for the 
exponent $\delta$ was obtained for any chaotic billiard with a rectangular 
bouncing-ball region \cite{BaeSchSti1997}. 
For the stadium billiard this bound agrees with $\delta=3/4$ from 
Ref.~\cite{Tan1997}. For the cosine billiard a bound of $\delta=9/10$ was 
obtained and $\delta\approx 0.87$ was observed numerically.

In this paper we relate the couplings between bouncing-ball modes 
and chaotic modes to the number $\Nbb(E)$ of bouncing-ball 
modes in a billiard.
The couplings give rise to decay rates $\gamma$, which describe the initial 
exponential decay $\sim$$\ue^{-\gamma t}$ of states concentrating on the 
marginally stable periodic orbits to the chaotic region of phase space.
In order to predict the decay rates $\gamma$ we employ the fictitious integrable 
system approach \cite{BaeKetLoeSch2008,BaeKetLoe2010}, which was previously 
used to determine regular-to-chaotic tunneling rates in systems with a mixed 
phase space \cite{BaeKetLoeSch2008,BaeKetLoeRobVidHoeKuhSto2008,
BaeKetLoeWieHen2009,LoeBaeKetSch2010,BaeKetLoe2010}.
We find a power-law decrease of the decay rates with increasing energy. 
We argue that the decreasing $\gamma$ imply the semiclassical existence of the 
bouncing-ball modes, complementing previous approaches 
\cite{OcoHel1988,Tan1997,Has2010}.
Furthermore, we use this prediction of decay rates to count the number 
of bouncing-ball modes $\Nbb(E)$. For the stadium billiard
we confirm the prediction of $\delta=3/4$ \cite{Tan1997,BaeSchSti1997}.
For the cosine billiard we derive the exponent $\delta=5/8=0.625$, which is well 
below the previously predicted upper bound $\delta=9/10$ \cite{BaeSchSti1997}.
Numerically we observe $\delta\approx 0.64$ using a larger energy range than in 
Ref.~\cite{BaeSchSti1997}, where $\delta\approx 0.87$ was found.

This paper is organized as follows. In Sec.~\ref{sec:dyntun} we review the 
fictitious integrable system approach, derive a prediction for the decay 
rates of bouncing-ball modes, and compare the results to numerically 
determined rates. As examples we consider the ergodic stadium billiard and 
the cosine billiard at parameters, where no regular motion is visible in 
phase space. In Sec.~\ref{sec:countbb} we use the prediction of decay rates 
to determine the number $\Nbb(E)$ of bouncing-ball modes and compare the 
results to the literature \cite{Tan1997,BaeSchSti1997}. 
A summary is given in Sec.~\ref{sec:summary}.


\section{Coupling of bouncing-ball modes}
\label{sec:dyntun}

For the analysis of the coupling of bouncing-ball modes to the chaotic sea
we use the analogy to systems with a mixed phase space. There the coupling
of regular and chaotic modes is caused by dynamical tunneling \cite{DavHel1981},
which quantum mechanically connects the classically separated regular and 
chaotic regions.
The billiards studied in this paper have no regular islands in phase space. 
However, there is a region surrounding the marginally stable bouncing-ball 
orbits which acts like a regular island, whose border is energy 
dependent such that its area semiclassically goes to zero \cite{Tan1997}.
For this situation we apply the fictitious integrable system approach 
\cite{BaeKetLoeSch2008,BaeKetLoe2010} to predict
the couplings of bouncing-ball modes to the chaotic sea.
Whether this coupling occurs due to dynamical tunneling as in mixed systems, 
or due to classically allowed transitions, e.g., through partial barriers, 
or some other mechanism is an open question that we do not address 
in this paper.

\subsection{The fictitious integrable system approach}
\label{sec:dyntun:fisa}

We first give a brief review of the fictitious integrable system approach 
\cite{BaeKetLoe2010}
which was previously used to determine regular-to-chaotic tunneling rates
in systems with a mixed phase space 
\cite{BaeKetLoeSch2008,BaeKetLoeRobVidHoeKuhSto2008,BaeKetLoeWieHen2009,
LoeBaeKetSch2010,BaeKetLoe2010}. 
Here it is applied in order to predict
couplings of bouncing-ball modes in chaotic billiards.
The main idea of the fictitious integrable system approach 
is the decomposition of Hilbert space into two parts, which 
correspond to the bouncing-ball modes (regular subspace) 
and to the remaining chaotic modes (chaotic subspace).
Such a decomposition can be found by introducing a fictitious integrable system 
$\Hreg$. This system has to be chosen such that its classical dynamics 
is integrable and contains the marginally stable bouncing-ball motion of the 
chaotic billiard. Quantum mechanically, the eigenstates $|\psiregmn\rangle$ of 
$\Hreg$, $\Hreg|\psiregmn\rangle = \Eregmn|\psiregmn\rangle$, 
closely resemble the bouncing-ball modes
of the chaotic billiard and are used as a basis for the regular subspace. 
The regular modes $|\psiregmn\rangle$ are characterized 
by two quantum numbers $m$ and $n$, where $m$ describes the quantization 
along the bouncing-ball orbits and $n$ perpendicular to them. They 
localize on quantizing tori of $\Hreg$, and decay beyond. 
This decay is the decisive property of $|\psiregmn\rangle$, which have no 
chaotic admixture, in contrast to bouncing-ball modes in a chaotic billiard.
Two choices for the system $\Hreg$ will be discussed in 
Sec.~\ref{sec:dyntun:int_app}.

Introducing a basis $|\psich\rangle$ in the chaotic subspace, the coupling 
matrix element of one regular mode $|\psiregmn\rangle$ and a chaotic 
mode $|\psich\rangle$ is given by
\begin{equation}
\label{eq:dyntun:fisa:vchbb}
 \vchmn = \langle\psich|H|\psiregmn\rangle.
\end{equation}
The disadvantage of these couplings $\vchmn$ is their dependence on the size 
of the chaotic region via the normalization of the chaotic modes 
$|\psich\rangle$.
A better suited quantity, that is not affected by such changes, is the decay
rate $\gamma$. Under variation of the billiard boundary far away from the 
rectangular bouncing-ball region it is unique for 
each regular state $|\psiregmn\rangle$. From the couplings $\vchmn$ the decay 
rate is obtained with Fermi's golden rule
\begin{equation}
\label{eq:dyntun:fisa:gamma}
 \gamma_{mn} = 2\pi \langle |\vchmn|^2 \rangle \rho_{\text{ch}}.
\end{equation}
Here we average over the modulus squared of the fluctuating 
coupling matrix elements $\vchmn$ of one particular regular mode and different 
chaotic modes of similar energy. The chaotic density of states is approximated
by the leading Weyl term $\rho_{\text{ch}} \approx A/(4\pi)$, 
in which $A$ denotes the area of the billiard.

\begin{figure*}[tb]
  \begin{center}
     \includegraphics[]{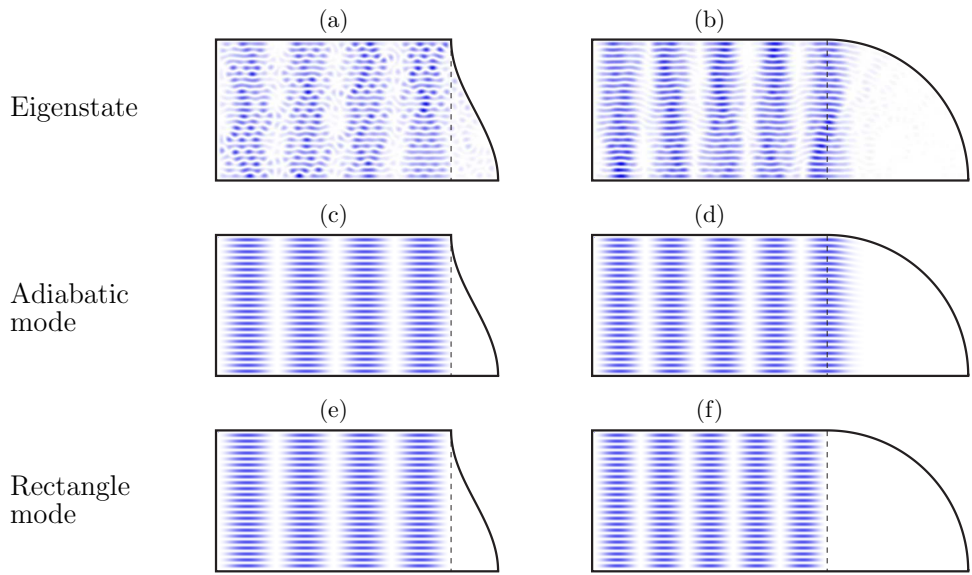}
     \caption{(Color online) Comparison of bouncing-ball modes of (a) the cosine 
              billiard with quantum number $(23,4)$ and (b)
              the stadium billiard with quantum number $(23,5)$ 
              to the corresponding adiabatic modes [(c), (d)] and to the 
              rectangle modes [(e), (f)]. 
              The adiabatic modes (c) and (d) closely resemble the bouncing-ball 
              modes (a) and (b), respectively.
              The rectangle mode (e) closely resembles the bouncing-ball 
              mode (a) of the cosine billiard, while for the stadium billiard 
              the rectangle mode (f) fails to reproduce the bouncing-ball 
              mode (b).}
     \label{fig:tunnel:quasimodes:comb}
  \end{center}
\end{figure*}

To illustrate the decay rates $\gamma$ one may consider a regular state 
$|\psiregmn\rangle$ concentrating close to the marginally stable bouncing-ball 
orbits which is coupled to a continuum of chaotic states. Its decay 
$\sim$$\ue^{-\gamma_{mn} t}$ is characterized by the rate $\gamma_{mn}$. 
For systems with a finite phase space this exponential decay occurs at most up 
to the Heisenberg time $\tau_H = 2\pi/\Delta$, where $\Delta$ is the mean 
level spacing. Alternatively, the decay rates are the inverse of the lifetimes 
of resonances in a corresponding open system, which can be obtained, e.g., 
by adding an absorbing region in the chaotic component of phase space far 
away from the bouncing-ball region.

\subsection{Integrable approximations}
\label{sec:dyntun:int_app}

In the following we predict decay rates $\gamma$ of bouncing-ball modes with 
Eqs.~\eqref{eq:dyntun:fisa:vchbb} and \eqref{eq:dyntun:fisa:gamma}.
For this we construct a fictitious integrable system $\Hreg$, whose eigenstates 
$|\psiregmn\rangle$ resemble the bouncing-ball modes of the chaotic 
billiard. Two approaches will be presented below.
The first uses adiabatic modes, which approximate the bouncing-ball modes quite 
well, but have to be evaluated numerically. The second approach uses  
modes of the rectangular billiard, which are given analytically, but 
turn out to be a good approximation for 
the bouncing-ball modes of the cosine billiard only.

\subsubsection{Adiabatic-mode approximation}
\label{sec:dyntun:ad}

In order to construct modes which approximate the bouncing-ball modes 
of a billiard we use the adiabatic separation ansatz \cite{BaiHosSteTay1985}
\begin{equation}
\label{eq:dyntun:ad:sep}
 \psiregmn(x,y) = \varphi_m(y;x)\chi_{mn}(x)
\end{equation}
with 
\begin{equation}
\label{eq:dyntun:ad:phi}
 \varphi_m(y;x) = \sqrt{\frac{2}{h(x)}} \sin\left(\frac{m\pi y}{h(x)}\right),
\end{equation}
where $h(x)$ is the height of the billiard at $x$ with $h(0) = h$. 
Equation~\eqref{eq:dyntun:ad:sep} is close to a separation ansatz but 
$\varphi_{m}$ weakly depends on $x$ in order to satisfy the Dirichlet boundary 
condition on $\partial \Omega$. 
The function $\varphi_{m}$ accounts for the quantization of the fast 
bouncing-ball motion in the $y$ direction. The slow motion in the $x$ 
direction is quantized by demanding that $\chi_{mn}$ fulfills the 1D 
Schr\"odinger equation \cite{BaiHosSteTay1985},
\begin{equation}
\label{eq:dyntun:ad:sglchi}
 -\chi_{mn}^{''}(x) + m^2V(x)\chi_{mn}(x) = e_{mn}\chi_{mn}(x),
\end{equation}
which we solve numerically.
Here the effective potential is $m^2V(x)$ with
\begin{equation}
\label{eq:dyntun:ad:pot}
 V(x) = \frac{\pi^2}{h(x)^2}-\frac{\pi^2}{h^2}
\end{equation}
and the eigenvalues are denoted by $e_{mn}$. 
From $e_{mn}$ one obtains approximate 
eigenvalues of the bouncing-ball modes $\Eregmn = m^2\pi^2/h^2 + e_{mn}$.
The ansatz, Eq.~\eqref{eq:dyntun:ad:sep}, exactly satisfies the Schr\"odinger 
equation inside the rectangular region of the billiard. Outside of the 
rectangular region this is only approximately true. There the functions 
$\chi_{mn}(x)$ will eventually decay exponentially, as 
the effective potential $m^2V(x)$ becomes very steep for large $m$.

The adiabatic modes show good agreement to the bouncing-ball modes of 
the stadium and the cosine billiard if $n \ll m$; see 
Figs.~\ref{fig:tunnel:quasimodes:comb}(a)-\ref{fig:tunnel:quasimodes:comb}(d). 
In particular the bouncing-ball modes are well reproduced in the region 
$x\approx w$, where the rectangular region ends.
Note that small deviations arise for modes with large $n$, in particular 
for the stadium billiard (not shown). However, these 
modes strongly couple to chaotic modes and thus do not contribute to the 
counting function in Sec.~\ref{sec:countbb}.

We now predict decay rates of bouncing-ball modes
using the adiabatic modes $|\psiregmn\rangle$.
Replacing the coupling matrix elements $\vchmn$ in 
Eq.~\eqref{eq:dyntun:fisa:gamma} by Eq.~\eqref{eq:dyntun:fisa:vchbb} we find
for the decay rates
\begin{eqnarray}
\label{eq:dyntun:fisa:gamma_inserted}
 \gamma_{mn} & = & \frac{2\pi\rho_{\text{ch}}}{N_\Delta} \langle\psiregmn | 
                    H \left(\sum_{\text{ch},\Delta E}|
                    \psich\rangle\langle\psich|\right) H|\psiregmn\rangle\\
\label{eq:dyntun:fisa:gamma_inserted_with_Pch}
             & = & \frac{2\pi\rho_{\text{ch}}}{N_\Delta} \Vert\PchdE H 
                   |\psiregmn\rangle\Vert^2.
\end{eqnarray}
To obtain Eq.~\eqref{eq:dyntun:fisa:gamma_inserted} we express 
the average in Eq.~\eqref{eq:dyntun:fisa:gamma} by a sum
over all chaotic states $|\psich\rangle$ in an energy 
interval $\Delta E$ around the energy $\Eregmn$ of the adiabatic
mode and $N_\Delta$ is the number of chaotic states in
this energy interval.
The term $\sum_{\text{ch},\Delta E} |\psich\rangle \langle\psich|$ in 
Eq.~\eqref{eq:dyntun:fisa:gamma_inserted} is equal to a projector 
$\PchdE$ onto the chaotic subspace in the considered energy interval.
We approximate this projector using the adiabatic modes $|\psiregmn\rangle$,
\begin{equation}
\label{eq:dyntun:fisa:gamma_pchde}
 \PchdE \approx \PdE \left(\opone - \sum_{m'\geq n'} 
        |\psiregmsns\rangle\langle\psiregmsns|\right).
\end{equation}
Here $\PdE$ projects onto the energy interval 
$[\Eregmn-\Delta E/2, \Eregmn+\Delta E/2]$. 
The choice $m'\geq n'$ for the sum ensures that $\PchdE$ projects onto the 
phase-space component with $k_x/h \geq k_y/w$, which excludes the bouncing-ball 
region but also parts of the surrounding chaotic region.
This choice will underestimate the decay rates slightly. However, a more 
precise projection would require an a priori knowledge of the energy-dependent
border between the bouncing-ball region and the chaotic region.
Using Eq.~\eqref{eq:dyntun:fisa:gamma_pchde} in 
Eq.~\eqref{eq:dyntun:fisa:gamma_inserted_with_Pch} the decay rates are 
determined by
\begin{equation}
\label{eq:dyntun:fisa:gamma2}
 \gamma_{mn} = \frac{A}{2N_\Delta} \left\Vert \PdE H|\psiregmn\rangle - 
               \hspace*{-0.3cm}\sum_{(m',n')\in\Delta E}\hspace*{-0.3cm} 
               \vmnmsns| \psiregmsns\rangle \right\Vert^2, 
\end{equation}
where 
\begin{equation}
\label{eq:dyntun:fisa:gamma2_vrr}
 \vmnmsns = \langle \psiregmsns| H |\psiregmn\rangle
\end{equation}
and $\rho_{\text{ch}}\approx A/(4\pi)$ has been used.
In order to numerically evaluate the term $\PdE H|\psiregmn\rangle$ 
in Eq.~\eqref{eq:dyntun:fisa:gamma2}
we expand the state $|\psiregmn\rangle$ in the eigenbasis $|\psi_l\rangle$ 
of $H$ in the energy interval $[\Eregmn-\Delta E/2, \Eregmn+\Delta E/2]$ such 
that $\PdE H|\psiregmn\rangle = \sum_{\Delta E} c_l E_l |\psi_l\rangle$, where 
$c_l=\langle\psi_{l}|\psiregmn\rangle$.
With this procedure Eq.~\eqref{eq:dyntun:fisa:gamma2} can be evaluated
numerically for all bouncing-ball modes of quantum number $(m,n)$. 
We find good agreement to numerically determined rates for the cosine and the 
stadium billiard; see Sec.~\ref{sec:dyntun:application}.

\subsubsection{Rectangle-mode approximation}
\label{sec:dyntun:rect}

While the approach for the determination of decay rates 
presented in the last section is applicable for any billiard with a 
rectangular bouncing-ball region, an analytical result is not available.
In order to find such an analytical result we now approximate 
the bouncing-ball modes by the eigenstates $|\psiregmn\rangle$ 
of a rectangular billiard and construct the chaotic states 
$|\psich\rangle$ entering in Eq.~\eqref{eq:dyntun:fisa:vchbb} by a 
random-wave model \cite{Ber1977}. As chaotic states $|\psich\rangle$ 
modeled in this way are not orthogonal to the regular states
$|\psiregmn\rangle$, Eq.~\eqref{eq:dyntun:fisa:vchbb} has to be modified 
(see Sec.~II~A in Ref.~\cite{BaeKetLoe2010}), which leads to the approximation
\begin{equation}
\label{eq:dyntun:fisa:vchbb2}
 \vchmn = \langle\psich|H-\Hreg|\psiregmn\rangle.
\end{equation}

The simplest approximation of a rectangular bouncing-ball region of width 
$w$ and height $h$  is given by the rectangular billiard of the same width and 
height. We now use this billiard as the fictitious integrable system $\Hreg$ in
Eq.~\eqref{eq:dyntun:fisa:vchbb2}. Its eigenstates in position representation 
are given analytically by
\begin{equation}
\label{eq:dyntun:rect:psibb}
\psiregmn(x,y) = \frac{2}{\sqrt{wh}} \sin\left(\frac{n\pi x}{w}\right)
                 \sin\left(\frac{m\pi y}{h}\right)
\end{equation}
with eigenenergies $\Eregmn = \pi^2 m^2 / h^2 + \pi^2 n^2 / w^2$.
Note that these rectangle modes $\psiregmn(x,y)$ are zero for $x\geq w$,
while the bouncing-ball modes of chaotic billiards will typically
extend into this region. Hence, the rectangle modes 
will only be an approximation of the true
eigenstates of the billiard. Figure~\ref{fig:tunnel:quasimodes:comb} shows that 
the rectangle modes are good approximations for the
bouncing-ball modes of the cosine billiard, which has a corner of angle 
$\pi/2$ at $x=w$, but not for the bouncing-ball modes of the stadium billiard, 
which substantially extend beyond $x=w$. 
As the fictitious integrable system approach relies on a good approximation 
of the bouncing-ball modes, the following prediction of decay rates 
is expected to apply only for billiards which have a corner 
of angle $\pi/2$ at $x=w$, like the cosine billiard. 

When evaluating Eq.~\eqref{eq:dyntun:fisa:vchbb2} 
an infinite potential difference 
arises between the Hamiltonian $H$ of the chaotic billiard
and $\Hreg$ of the rectangle, $H-\Hreg= -\infty$, for $x\geq w$. At the same
time for the rectangle modes $\psiregmn(x,y)=0$ holds in that region, 
which leads to an undefined product $-\infty \cdot 0$. Similar to the approach
presented for the mushroom billiard in 
Refs.~\cite{BaeKetLoeRobVidHoeKuhSto2008,BaeKetLoe2010} we circumvent this
problem by introducing a rectangular billiard which is extended by a finite 
potential $W$ at $x\geq w$ for which in the end the limit $W\to\infty$ is taken.
This leads to
\begin{equation}
\label{eq:dyntun:rect:vchmnderiv_0}
 \vchmn = \Int_{0}^{h}\ud y\,\psich(x=w,y)\,\partial_x\psiregmn(x=w,y).
\end{equation}

For completeness we now give the derivation of 
Eq.~\eqref{eq:dyntun:rect:vchmnderiv_0} following 
Refs.~\cite{BaeKetLoeRobVidHoeKuhSto2008,BaeKetLoe2010}:
We first introduce a rectangular billiard which is extended by a finite 
potential $W$ for $x\geq w$,
\begin{eqnarray}
\label{eq:dyntun:rect:HregW}
 \Hreg^W & = & p_{x}^2 + p_{y}^{2} + V(x,y)\\
 V(x,y) & = & \left\{\begin{array}{ll} 0& \quad \text{for } 0<x<w,\; 0<y<h\\ 
                              W& \quad \text{for } x\geq w,\; 0<y<h\\
                              \infty& \quad \text{otherwise}\end{array}\right. 
\end{eqnarray}
and consider the limit $W\to\infty$ in which the original rectangular billiard 
is recovered. For finite $W$ the regular eigenfunctions $\psiregW^{mn}(x,y)$ of
$\Hreg^W$ decay into the region $x\geq w$, which is described by 
\begin{eqnarray}
 \psiregW^{mn}(x,y) = \psiregW^{mn}(x=w,y)\ue^{-\lambda (x-w)}.
\end{eqnarray}
Here $\lambda$ depends on $W$ via the Schr\"odinger equation as
\begin{eqnarray}
\label{eq:dyntun:rect:HregW_SGL}
 -\lambda^2 + W = E_{mn}^{W}.
\end{eqnarray}
Since the derivative of the regular eigenfunctions $\psiregW^{mn}$ has to 
be continuous at $x=w$ we obtain
\begin{eqnarray}
 \partial_x \psiregW^{mn}(x=w,y) = -\lambda \,\psiregW^{mn}(x=w,y).
\end{eqnarray}
This can be used for rewriting Eq.~\eqref{eq:dyntun:fisa:vchbb2} for the
coupling matrix elements 
\begin{eqnarray}
\label{eq:dyntun:rect:v_final1}
 \vchmn & \hspace*{-0.05cm}= & \hspace*{-0.1cm}\lim_{W\to\infty} 
          \hspace*{-0.1cm} \Int_{w}^{w+b}
          \hspace*{-0.1cm}\ud x  \hspace*{-0.15cm} \Int_{0}^{h(x)}
          \hspace*{-0.15cm} \ud y\, \psich(x,y)
           (-W)\psiregW^{mn}(x,y)\quad\;\;\\
\label{eq:dyntun:rect:v_final1b}
        & \hspace*{-2.2cm}= & \hspace*{-1.3cm}\lim_{W\to\infty} 
          \hspace*{-0.15cm}
          \Int_{w}^{w+b}\hspace*{-0.15cm}\ud x  \hspace*{-0.2cm} \Int_{0}^{h(x)}
          \hspace*{-0.2cm} \ud y\, \psich(x,y)\frac{W}{\lambda}
          \ue^{-\lambda (x-w)}\partial_{x}\psiregW^{mn}(w,y).
\end{eqnarray}
The term 
\begin{eqnarray}
 \frac{W}{\lambda}\ue^{-\lambda (x-w)} = \frac{W}{\sqrt{W-E_{mn}^{W}}}
                  \ue^{-\sqrt{W-E_{mn}^{W}} (x-w)}
\end{eqnarray}
reduces in the limit $W\to\infty$ to $2\delta(x-w)$ for $x \geq w$,
where we used Eq.~\eqref{eq:dyntun:rect:HregW_SGL} and that $E_{mn}^{W}$ 
remains bounded. This finally leads to Eq.~\eqref{eq:dyntun:rect:vchmnderiv_0}.

Evaluating the coupling matrix elements, 
Eq.~\eqref{eq:dyntun:rect:vchmnderiv_0}, 
for the rectangle modes, Eq.~\eqref{eq:dyntun:rect:psibb}, gives
\begin{equation}
\label{eq:dyntun:rect:vchmnderiv}
 \vchmn = (-1)^n\frac{2}{\sqrt{wh}}\frac{n\pi}{w}\Int_{0}^{h}\ud y\,
          \psich(x=w,y)\,\sin\left(\frac{m \pi y}{h}\right).
\end{equation}
Therefore the decay rates $\gamma_{mn}$, Eq.~\eqref{eq:dyntun:fisa:gamma}, 
scale with $n^2$.

\begin{figure}[tb]
  \begin{center}
     \includegraphics[]{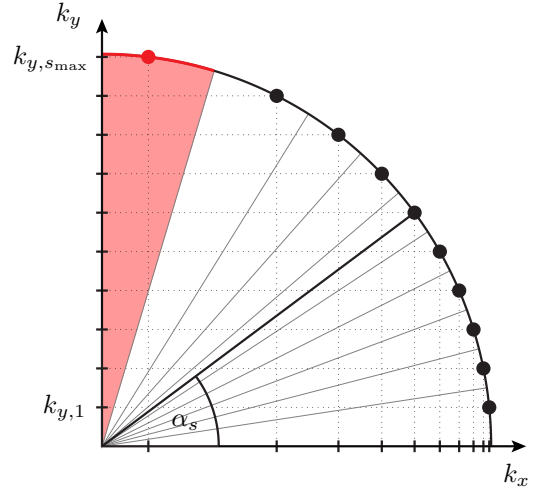}
     \caption{(Color online) Illustration of the discrete directions 
              contributing to the random-wave model, 
              Eq.~\eqref{eq:dyntun:rect:psich}, in $k$ space
              at energy $\Eregmn$ (black quarter circle) with $m=10$ and $n=2$.
              While the wave numbers $k_{y,s}$ are equispaced,
              the wave numbers $k_{x,s}$ as well as the 
              angles $\alpha_s$ are not equispaced. 
	      The angular regions around each $\alpha_s$ are marked by gray 
              lines and the one for $\alpha_{s_{\text{max}}}$ is shaded.
              }
     \label{fig:tunnel:rect:rwm}
  \end{center}
\end{figure}

In order to find the scaling of the decay rates with $m$ we 
employ a random-wave model \cite{Ber1977} for the chaotic states $\psich(x,y)$ 
entering in Eq.~\eqref{eq:dyntun:rect:vchmnderiv}. Here it is given by
\begin{eqnarray}
\label{eq:dyntun:rect:psich}
 \psich(x,y) = \frac{2}{\sqrt{A}}\sum_{s=1}^{s_{\text{max}}} 
               c_s \sin\left(\frac{s\pi y}{h}\right) 
               \sin(k_{x,s} x+\varphi_s)\;\;
\end{eqnarray} 
and accounts for the Dirichlet boundary conditions at $y=0$ and $y=h$,
but ignores the corner at $x=w$. 
Here the $c_s$ are Gaussian-distributed random variables with mean zero and 
variance $\sigma_s=\langle c_{s}^{2}\rangle$. The phases $\varphi_s$ are 
uniformly distributed in $[0,2\pi)$, $k_{x,s}=\sqrt{\Eregmn-\pi^2 s^2/h^2}$, 
and $s_{\text{max}} = \lfloor (h/\pi)\sqrt{\Eregmn} \rfloor$. 
Furthermore we define $k_{y,s}=\pi s/h$ and the angles
$\alpha_s = \arctan(k_{y,s}/k_{x,s})$. These angles are not equidistributed 
on $[0,\pi/2]$; see Fig.~\ref{fig:tunnel:rect:rwm}. To approximate an 
equidistribution of directions we compensate this
by appropriate weights $\sigma_s$, with 
$\sum_{s=1}^{s_{\text{max}}}\sigma_s =1$.
A natural choice is given by assigning to each $\sigma_s$ half of the 
size of the adjacent angular intervals
\begin{equation}
\label{eq:dyntun:rect:sigma}
 \sigma_s = \frac{1}{\pi/2} \left\{ 
       \begin{array}{ll}
         (\alpha_{s+1}-\alpha_{s-1})/2\;, & 1<s<s_{\text{max}}   \\
         \pi/2-(\alpha_{s_{\text{max}}} + \alpha_{s_{\text{max}}-1})/2\;, 
           & s=s_{\text{max}}\\
         (\alpha_1 + \alpha_2)/2\;, & s=1,                
       \end{array}
\right.
\end{equation}
see Fig.~\ref{fig:tunnel:rect:rwm}.

Using Eq.~\eqref{eq:dyntun:rect:psich} in Eq.~\eqref{eq:dyntun:rect:vchmnderiv}
the $y$ integration over the two sine functions gives $\delta_{sm} h/2$
such that 
\begin{eqnarray}
\label{eq:dyntun:rect:vchmnderiv2}
 \vchmn = (-1)^n\frac{2}{\sqrt{wh}}\frac{n\pi}{w}\frac{h}{\sqrt{A}} 
          c_m \sin(k_{x,m} w+\varphi_m).\;\;\;
\end{eqnarray}
We now determine the decay rates with Eq.~\eqref{eq:dyntun:fisa:gamma} 
using an ensemble average of the modulus squared of the coupling matrix elements
given by Eq.~\eqref{eq:dyntun:rect:vchmnderiv2}. This results in
\begin{eqnarray}
\label{eq:dyntun:rect:gammares1}
 \gamma_{mn} = \frac{\pi^2 h}{w^{3}} n^2\sigma_m
\end{eqnarray}
which contains the weight $\sigma_m$.
In order to find the weight $\sigma_m$ from Eq.~\eqref{eq:dyntun:rect:sigma}
we have to evaluate $s_{\text{max}}$. 
One finds $s_{\text{max}}=m$ for small enough $n<\sqrt{2(m+1/2)}l/h$
following from $(h/\pi)\sqrt{\Eregmn} < m+1$.
We now perform a first-order expansion of the angles $\alpha_m$ in 
Eq.~\eqref{eq:dyntun:rect:sigma} for large $m$ 
and restrict to even smaller $n\ll\sqrt{m}$. We obtain 
\begin{equation}
\label{eq:dyntun:rect:sigma2}      
    \sigma_m \approx \frac{\sqrt{2}}{\pi} \frac{1}{\sqrt{m}}
\end{equation}
leading to our final result
\begin{eqnarray}
\label{eq:dyntun:rect:gammares2}
 \gamma_{mn} = \frac{\sqrt{2}\pi h}{w^{3}} \frac{n^2}{\sqrt{m}},
\end{eqnarray}
which predicts the decay rates of bouncing-ball modes characterized by the 
quantum numbers $m$ and $n$ to the chaotic sea for $n\ll\sqrt{m}$. 
For increasing $m$ the decay rates decrease like a square root while for 
increasing $n$ they increase quadratically. 

The power-law decrease of the decay rates for constant $n$ 
and increasing $m$ is in contrast to the exponential decrease found for 
direct regular-to-chaotic tunneling rates in mixed systems, 
as in the mushroom billiard \cite{BarBet2007,BaeKetLoeRobVidHoeKuhSto2008} 
and in quantum maps 
\cite{HanOttAnt1984,PodNar2003,SheFisGuaReb2006,BaeKetLoeSch2008}.
The decreasing decay rates for fixed $n$ lead to decreasing couplings 
$\vchmn$, see Eq.~\eqref{eq:dyntun:fisa:gamma}, between the bouncing-ball 
modes and chaotic modes compared to the mean level spacing $\Delta$. 
This implies for large enough $m$ the semiclassical existence 
of the bouncing-ball mode $(m,n)$.
Note that the prediction, Eq.~\eqref{eq:dyntun:rect:gammares2}, only depends on
the rectangular bouncing-ball region but not on the other
regions of the considered billiard. Hence, it is the same for each billiard 
studied in this paper. However, it can only be applied if the rectangle modes 
closely resemble the bouncing-ball modes. 
This is the case for the cosine billiard but not for the stadium billiard; 
see Sec.~\ref{sec:dyntun:application} for the comparison of 
Eq.~\eqref{eq:dyntun:rect:gammares2} to numerically determined rates.

\subsection{Applications}
\label{sec:dyntun:application}

We now evaluate Eqs.~\eqref{eq:dyntun:fisa:gamma2} and 
\eqref{eq:dyntun:rect:gammares2} for the decay rates of 
bouncing-ball modes, as discussed in Sec.~\ref{sec:dyntun:int_app}, 
and compare the results to numerically determined decay 
rates for specific example systems. We choose the desymmetrized cosine 
billiard with $w=1$, $h=0.6$, $b=0.2$, $h(x)=h\arccos(2(x-w)/b-1)/\pi$ for 
$x>w$, and label this billiard by $\cosb$. Furthermore we consider the 
desymmetrized stadium billiard with
$w=1$, $h=0.6$, $h(x)=\sqrt{h^2-(x-w)^2}$ for $x>w$, and label this 
billiard by $\stadb$.

For the evaluation of Eq.~\eqref{eq:dyntun:fisa:gamma2} the energy interval 
$\Delta E$ has to be specified. We choose $\Delta E \approx 100\Delta$ with mean 
level spacing $\Delta$.
Note that for small energy intervals $\Delta E<20\Delta$ the results of 
Eq.~\eqref{eq:dyntun:fisa:gamma2} show strong fluctuations, 
as the adiabatic states cannot be properly expanded in the eigenbasis of $H$. 
For $\Delta E>20\Delta$, however, the results show only small changes. 

In order to calculate decay rates numerically we determine the spectrum
of the stadium and the cosine billiard, using the improved method of 
particular solutions \cite{BetTre2005}, under variation of parts of the 
billiard boundary: For the cosine billiard we vary the width of the cosine 
part $b$ around $b=0.2$ and for the stadium we vary $w$ around $w=1$.
For small variations of these parameters the eigenvalues of the bouncing-ball 
modes remain almost unaffected while the eigenvalues of the chaotic states show 
strong variations, due to the changing density of chaotic states.
Analyzing up to $30$ avoided crossings 
$\Delta E_{\text{ch},mn}= 2\vchmn$ of a given bouncing-ball mode $\psi_{mn}$
with chaotic states we deduce the decay rate from Fermi's golden rule
\cite{BaeKetLoeRobVidHoeKuhSto2008,BaeKetLoe2010},
\begin{equation}
\label{eq:dyntun:rect:gammanum}
 \gamma_{mn} = \frac{1}{8} \langle |\Delta E_{\text{ch},mn}|^2 A\rangle.
\end{equation}
Here we average over all numerically determined widths
$\Delta E_{\text{ch},mn}$ and the corresponding billiard areas $A$.

\begin{figure}[tb]
  \begin{center}
     \includegraphics[width=85mm]{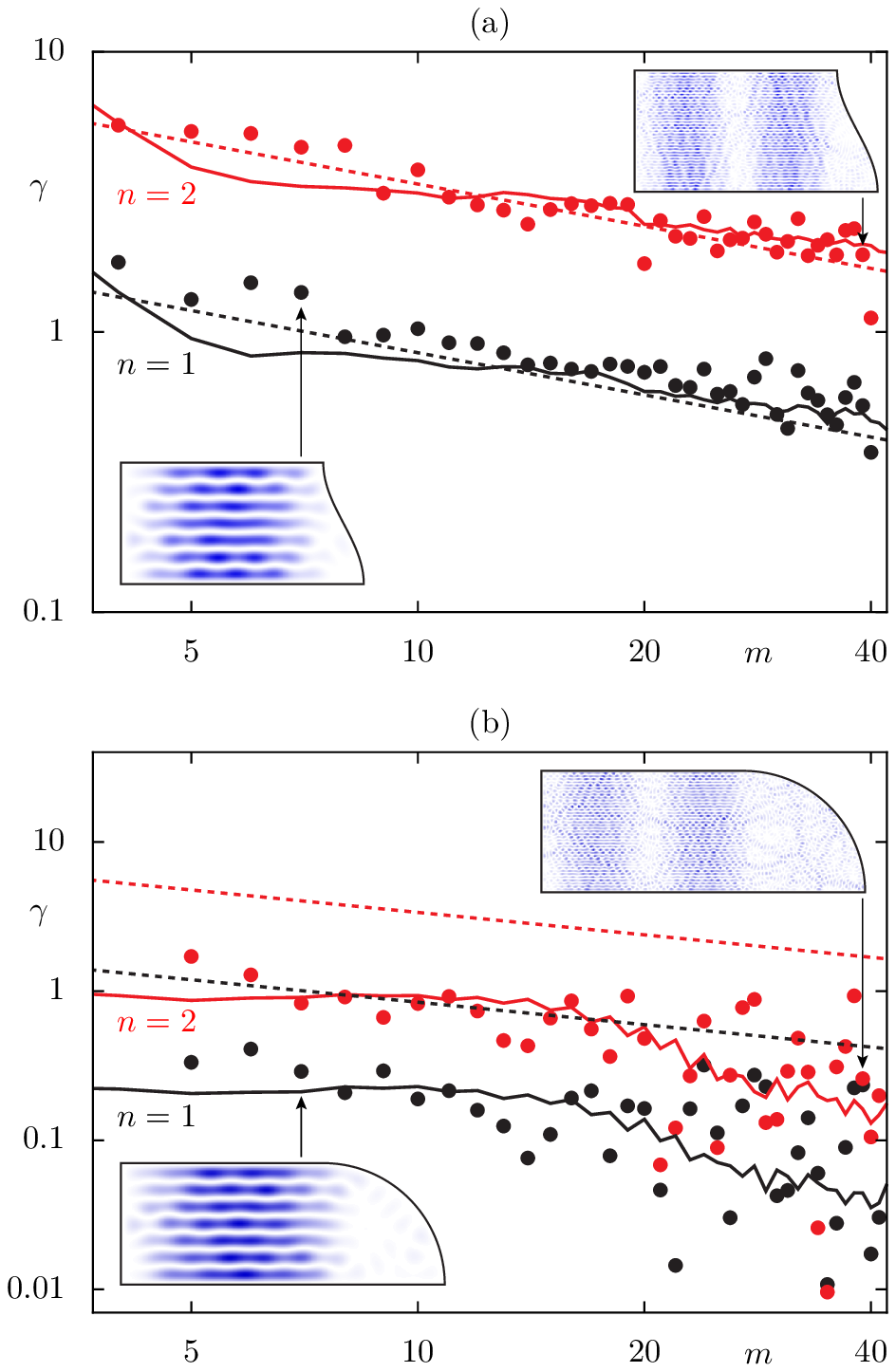}
     \caption{(Color online) Decay rates of bouncing-ball modes for (a) the 
              cosine billiard $\cosb$ and (b) the stadium billiard $\stadb$. 
              For the quantum numbers $n=1,2$ and increasing $m$ 
              we compare numerical rates (dots) to the results of 
              Eqs.~\eqref{eq:dyntun:fisa:gamma2} (solid lines) 
              and \eqref{eq:dyntun:rect:gammares2} (dashed lines)
              on a double-logarithmic scale.
              The insets in (a) and (b) show the bouncing-ball modes of 
              quantum number $(7,1)$ and $(39,2)$.}
     \label{fig:tunnel:comb}
  \end{center}
\end{figure}

In Fig.~\ref{fig:tunnel:comb}(a) we show for the cosine billiard $\cosb$
numerical rates (dots) compared with the prediction of 
Eq.~\eqref{eq:dyntun:fisa:gamma2}, using the adiabatic modes (solid lines),
and Eq.~\eqref{eq:dyntun:rect:gammares2}, 
using the rectangle modes (dashed lines).
We consider bouncing-ball modes with horizontal quantum number $n=1,2$ and 
increasing vertical quantum number $m$. We find very good agreement with 
deviations smaller than a factor of two for both predictions.
The numerical rates and the results of Eq.~\eqref{eq:dyntun:fisa:gamma2} 
follow the power law $\gamma \sim m^{-1/2}$ predicted by 
Eq.~\eqref{eq:dyntun:rect:gammares2}.
This confirms that for the cosine billiard the adiabatic modes as well as the 
rectangle modes sufficiently well approximate the bouncing-ball modes, 
as seen in Figs.~\ref{fig:tunnel:quasimodes:comb}(a), 
\ref{fig:tunnel:quasimodes:comb}(c), and \ref{fig:tunnel:quasimodes:comb}(e).
Note that also for the desymmetrized Sinai billiard \cite{Sin1970} both 
predictions agree with numerical rates (not shown). It also has
a corner of angle $\pi/2$ at $x=w$, such that the rectangle and the 
adiabatic modes closely resemble its bouncing-ball modes.

\begin{figure}[tb]
  \begin{center}
     \includegraphics[width=85mm]{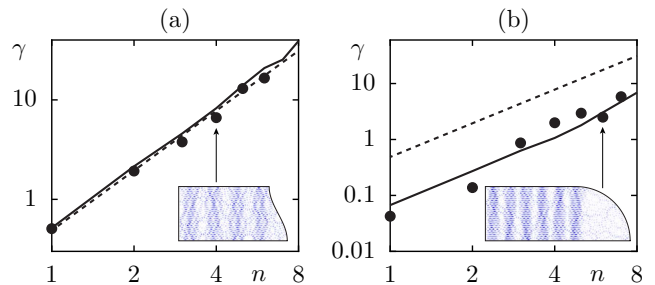}
     \caption{(Color online) Decay rates of bouncing-ball modes for (a) the 
              cosine billiard $\cosb$ and (b) the stadium billiard $\stadb$. 
              For the quantum numbers
              $m=30$ and $n=1,2,\dots$, we compare numerical rates (dots) to 
              the results of Eqs.~\eqref{eq:dyntun:fisa:gamma2} (solid lines) 
              and \eqref{eq:dyntun:rect:gammares2} (dashed lines) 
              on a double-logarithmic scale. The insets show the bouncing-ball 
              modes of quantum number (a) $(30,4)$ and (b) $(30,6)$.}
     \label{fig:tunnel:comb_var_n}
  \end{center}
\end{figure}

Figure~\ref{fig:tunnel:comb}(b) shows the decay rates for the stadium 
billiard $\stadb$ for $n=1,2$ and increasing $m$. We compare the numerical 
rates (dots) to the prediction of Eq.~\eqref{eq:dyntun:fisa:gamma2} 
(solid lines), and Eq.~\eqref{eq:dyntun:rect:gammares2} (dashed lines).  
Here, the average behavior of the fluctuating numerical rates agrees with the 
prediction of Eq.~\eqref{eq:dyntun:fisa:gamma2}. 
As for the cosine billiard this prediction seems to follow a power-law for 
large $m$, however, with an exponent smaller than $-1/2$, close to $-1$. 
As expected, Eq.~\eqref{eq:dyntun:rect:gammares2}, based on the rectangle modes, 
does not reproduce the numerical rates, showing deviations by a factor 
of $10$. These deviations arise as the rectangle modes do not agree well enough 
with the bouncing-ball modes of the stadium billiard, in contrast to the 
adiabatic modes used in Eq.~\eqref{eq:dyntun:fisa:gamma2}; 
see Figs.~\ref{fig:tunnel:quasimodes:comb}(b), 
\ref{fig:tunnel:quasimodes:comb}(d), and \ref{fig:tunnel:quasimodes:comb}(f).
Note that the small fluctuations in the prediction arise due to the 
numerical evaluation of Eq.~\eqref{eq:dyntun:fisa:gamma2} and might be related 
to the chosen projector $\PchdE$.
The large fluctuations in the numerically determined rates could be caused by 
additional couplings of the bouncing-ball modes to scars 
concentrating on unstable periodic orbits, which is left for future studies.

In Fig.~\ref{fig:tunnel:comb_var_n} we show numerically determined decay rates
and the two predictions, Eqs.~\eqref{eq:dyntun:fisa:gamma2} and 
\eqref{eq:dyntun:rect:gammares2}, for bouncing-ball modes with fixed 
quantum number $m=30$ and increasing $n$. 
For both the cosine and the stadium billiard we find good agreement between the
numerical rates and Eq.~\eqref{eq:dyntun:fisa:gamma2} using the adiabatic modes.
Equation~\eqref{eq:dyntun:rect:gammares2}, based on the rectangle modes, 
is valid only for the cosine billiard as discussed before. 
For increasing $n$ the decay rates increase almost 
quadratically, $\gamma \sim n^2$, for both of the billiards, 
as found in Eq.~\eqref{eq:dyntun:rect:gammares2}.


\section{Counting bouncing-ball modes}
\label{sec:countbb}    

We now study and predict the number of bouncing-ball modes $\Nbb(E)$ 
up to energy $E$ for a billiard with a rectangular region. 
Due to the quantum ergodicity theorem \cite{Shn1974} in an 
ergodic billiard the ratio of $\Nbb(E)$ to the total number of eigenstates 
$N(E)$ goes to zero in the semiclassical limit, $\Nbb(E)/N(E) \to 0$. 
In addition the number of bouncing-ball modes should increase at least as 
$\sqrt{E}$, which gives the number of states concentrating on a one-dimensional 
line. Hence, one expects for the number of bouncing-ball modes 
\begin{equation}
\label{eq:count:Nbb}
 \Nbb(E) = \alpha \cdot E^{\delta} + \dots\;
\end{equation}
with exponent $1/2 \leq \delta < 1$ and prefactor $\alpha$. 
Here we neglect higher order terms which
contain powers of the energy smaller than $\delta$.

Note that in Ref.~\cite{BaeKetLoeSch2011} the boundary contribution 
$\sim$$\sqrt{E}$ of Eq.~\eqref{eq:count:Nbb} has been determined for subsets 
of eigenstates in billiards. For the bouncing-ball modes in the billiards 
considered in this paper this boundary contribution suggests the additional term 
$-h/(4\pi)\sqrt{E}$. As we are not dealing with a regular region around a fixed 
point but an energy dependent region around the line of marginally stable 
bouncing-ball orbits, we expect the boundary contribution $-2h/(4\pi)\sqrt{E}$.
However, it will not be considered in the following, as we are interested in the 
leading term only.

\subsection{Counting using overlap with adiabatic modes}
\label{sec:countbb:over} 

First we count the number of bouncing-ball modes $\Nbb(E)$ 
in the cosine and the stadium billiard numerically.
For this purpose we calculate their first $\approx$$3000$ eigenstates 
$|\psi_l\rangle$ and eigenvalues $E_l$  
using the improved method of particular solutions \cite{BetTre2005}. 
We then determine the overlap 
\begin{equation}
 W = |\langle\psi_l|\psiregmn\rangle|^2
\end{equation}
of each eigenstate $|\psi_l\rangle$ with the adiabatic 
approximations $|\psiregmn\rangle$ of the bouncing-ball modes. If 
$|\psi_l\rangle$ has an overlap $W>1/2$ with an adiabatic mode with $m \geq n$
we consider $|\psi_l\rangle$ as a bouncing-ball mode of the system. 
Counting the number of these modes gives a numerical estimate of $\Nbb(E)$.

In Figs.~\ref{fig:count:comb:gamma}(a) and \ref{fig:count:comb:gamma}(b) the
results (black staircase functions) for the cosine and the stadium billiard are 
shown, respectively, on a double-logarithmic scale. For large energies the data 
follow in both cases a power law for the number of bouncing-ball modes,
as expected from Eq.~\eqref{eq:count:Nbb}. 
Note that varying the cut off $W=1/2$ in $[0.5,0.6]$ changes the 
prefactors $\alpha$ but not the exponents $\delta$ (not shown).   

Specifically for the stadium billiard one expects $\delta=3/4$ 
\cite{Tan1997,BaeSchSti1997}. In Fig.~\ref{fig:count:comb:gamma}(b)
we show the function $\Nbb(E) = 0.13 \cdot E^{3/4}$ (red dashed line),
which on a double-logarithmic scale has the same slope as the numerical data.
In the inset we fit an exponent 
$\delta \approx 0.73$ to the data which is close to $\delta=3/4$.

For the cosine billiard $\delta=9/10$ was obtained as an upper bound  
and $\delta\approx 0.87$ was observed numerically \cite{BaeSchSti1997}. 
However, in Fig.~\ref{fig:count:comb:gamma}(a) the 
data are best fitted by an exponent $\delta \approx 0.64$  
(red dashed line in the inset).
In order to resolve this contradiction we also studied the cosine billiard 
used in Ref.~\cite{BaeSchSti1997} with parameters 
$h=1$, $w=2$, and $b=1$ (not shown). 
In Ref.~\cite{BaeSchSti1997} numerical data for $\Nbb(E)$ is presented 
which is well described by $\Nbb(E) = \alpha E^\delta + \beta$ with an 
exponent $\delta\approx0.87$ and a constant offset $\beta$. 
For the identification of the bouncing-ball modes 
a visual selection is performed and states up to energy $E=8400$ are considered.
If we use the overlap criterion with $W>1/2$ for this cosine billiard and 
consider states up to energy $E=20000$ we obtain an exponent 
$\delta \approx 0.64$. 
Also a visual selection over this larger energy range gives an exponent 
$\delta\approx 0.7$. Hence, we conclude that the smaller energy interval 
used in Ref.~\cite{BaeSchSti1997} is not sufficient to find the asymptotic 
scaling exponent.

\subsection{Counting using decay rates}
\label{sec:countbb:rates}

In Refs.~\cite{BaeKetMon2005,BaeKetMon2007} the criterion 
$\gamma_{mn} < 1/\tau_H$ has been found to describe the 
existence of regular states for systems with a mixed phase space. 
It compares the decay rates $\gamma_{mn}$ 
to the Heisenberg time $\tau_H=h/\Delta=2\pi/\Delta$. 
Here we apply a similar criterion to predict the number of bouncing-ball 
modes in a billiard, using the results \eqref{eq:dyntun:fisa:gamma2} and 
\eqref{eq:dyntun:rect:gammares2} for the decay rates derived in 
Sec.~\ref{sec:dyntun}. 

\begin{figure}[tb]
  \begin{center}
     \includegraphics[width=85mm]{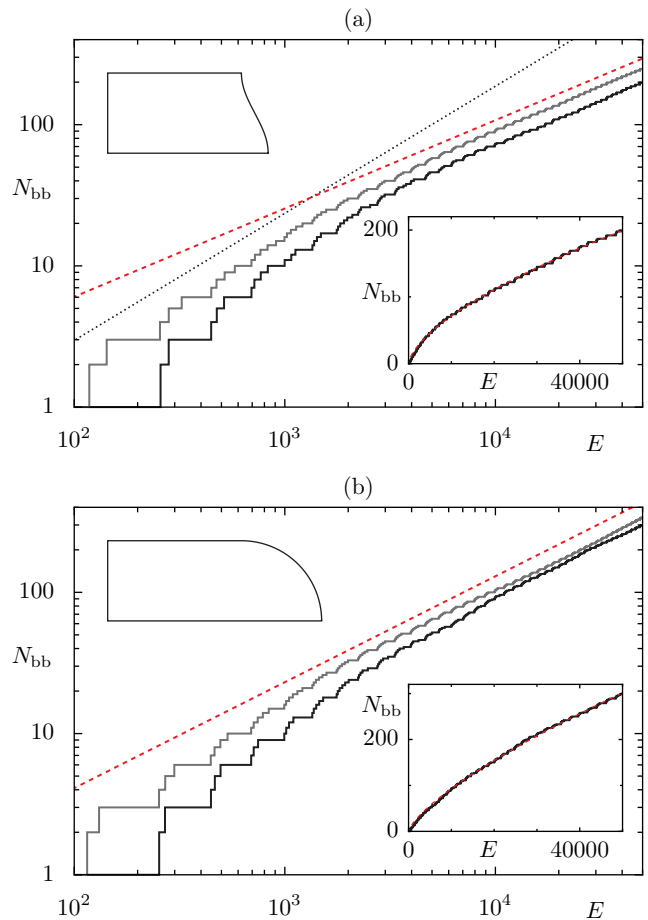}
     \caption{(Color online) Number of bouncing-ball modes $\Nbb(E)$ for (a) 
              the cosine billiard $\cosb$ and (b) the stadium billiard $\stadb$.
              We compare the results using the overlap criterion 
              (black staircase functions) and the prediction using 
              the decay rates (gray staircase functions) to power laws
              with exponent (a) $\delta=5/8$ and (b) $\delta=3/4$ 
              (red dashed lines) on a double-logarithmic scale. In (a) 
              we show in addition a power law with 
              exponent $\delta=9/10$ (dotted line) \cite{BaeSchSti1997}.
              The insets show the results using the overlap criterion on 
              a linear scale compared to
              Eq.~\eqref{eq:count:Nbb} fitted for $E>10000$ (red dashed lines). 
              We find $\delta=0.64$, $\alpha=0.196$ in (a) and $\delta=0.73$, 
              $\alpha=0.112$ in (b).}
     \label{fig:count:comb:gamma}
  \end{center}
\end{figure}

From Eq.~\eqref{eq:dyntun:fisa:gamma} we obtain the average coupling $\vmn$
of a bouncing-ball mode to the chaotic sea,
\begin{equation}
\label{eq:count:rates_couplings}
 \vmn = \sqrt{\frac{\gamma_{mn}}{2\pi\rho_{\text{ch}}}}.
\end{equation}
If this coupling $\vmn$ is much smaller than $\Delta$, on average the 
bouncing-ball mode weakly couples to the chaotic modes such that it is visible.  
If the coupling is much larger than $\Delta$, on average the bouncing-ball 
mode couples strongly to many chaotic modes such that it is not visible.
In between these two limiting cases there is a smooth transition, which was 
studied in Ref.~\cite{BaeKetMon2007} for systems with a mixed phase space. 
In order to count the number of bouncing-ball modes we approximate this smooth
transition by a sharp condition.
The criterion $\gamma_{mn} < 1/\tau_H$ would lead to the condition
$\vmn<\Delta/(2\pi)$. It is very strict in the sense that it only allows for 
very small couplings between the bouncing-ball modes and the chaotic modes.
Approximately in the middle of the transition we find the condition
\begin{equation}
\label{eq:count:rates_crit_coupling}
 \vmn < \frac{\Delta}{2}
\end{equation}
which we use in the following for counting the number of bouncing-ball modes. 

We now calculate the average coupling $\vmn$ for all bouncing-ball modes
up to energy $E$ with Eq.~\eqref{eq:count:rates_couplings}, using the decay 
rates $\gamma_{mn}$ from Eq.~\eqref{eq:dyntun:fisa:gamma2}.
We then count the number of those modes which fulfill 
the condition \eqref{eq:count:rates_crit_coupling}. 
The results are shown in Fig.~\ref{fig:count:comb:gamma}(a)
for the cosine billiard and in Fig.~\ref{fig:count:comb:gamma}(b)
for the stadium billiard as gray staircase functions.
For the stadium billiard we obtain the exponent $\delta\approx 0.76$  
close to $\delta=3/4$ \cite{Tan1997,BaeSchSti1997} 
and for the cosine billiard we find $\delta\approx 0.63$.
Note that changing the condition \eqref{eq:count:rates_crit_coupling} 
to $\vmn < \Delta/(2\pi)$ 
changes the prefactor $\alpha$ but not the exponent $\delta$
for the examples we considered (not shown).  
 
In order to obtain an analytical prediction of $\Nbb(E)$ for the cosine 
billiard we use the result for the decay rates $\gamma_{mn}$, 
Eq.~\eqref{eq:dyntun:rect:gammares2}. 
Together with Eq.~\eqref{eq:count:rates_couplings}, condition
\eqref{eq:count:rates_crit_coupling} is fulfilled by all bouncing-ball modes 
with 
\begin{equation}
\label{eq:count:gamma:nmrel}
 n < \sqrt{\frac{\pi\Delta}{2}}\sqrt{\frac{w^3}{\sqrt{2}\pi h}} m^{\frac{1}{4}}.
\end{equation}
For large $m$ this complies with the restriction $n\ll\sqrt{m}$ used 
in the derivation of Eq.~\eqref{eq:dyntun:rect:gammares2}.
For the determination of $\Nbb(E)$ one now has to count all states of quantum 
number $(m,n)$ for which Eq.~\eqref{eq:count:gamma:nmrel} is fulfilled and 
$\Eregmn = \pi^2m^2/h^2+\pi^2n^2/w^2 < E$. This sum can be approximated by an 
integral, where we integrate the right-hand side 
of Eq.~\eqref{eq:count:gamma:nmrel}
over $m$ in the interval $[0,m_{\text{max}}]$ and approximate 
$m_{\text{max}} \approx \sqrt{E}h/\pi$. This finally gives for the number of 
bouncing-ball modes 
\begin{eqnarray}
\label{eq:count:gamma:Nbbanal}
\Nbb(E) \approx \alpha\cdot E^{\frac{5}{8}}.
\end{eqnarray}
with the constant 
$\alpha=4w^{3/2}h^{3/4}\sqrt{\pi\Delta/2}/(5\cdot2^{1/4} \pi^{7/4})$.
Equation~\eqref{eq:count:gamma:Nbbanal} gives the exponent $\delta=5/8$ 
[red dashed line in Fig.~\ref{fig:count:comb:gamma}(a)]. It is close to
the exponents $\delta\approx 0.64$ and $\delta\approx 0.63$ 
which were obtained using the overlap criterion and 
the decay rates from Eq.~\eqref{eq:dyntun:fisa:gamma2}, respectively.
Even the prefactor $\alpha \approx 0.34$ agrees roughly with the fitted value 
$\alpha \approx 0.2$.
Note that Eq.~\eqref{eq:count:gamma:Nbbanal} is only valid for billiards
with a confining corner at $x=w$, as in the cosine billiard, because otherwise 
Eq.~\eqref{eq:dyntun:rect:gammares2} for the decay rates cannot be applied,
as in the case of the stadium billiard.


\section{Summary}
\label{sec:summary}

In this paper we study the couplings of bouncing-ball modes to chaotic 
modes in two-dimensional billiards and use these couplings to count the number
of bouncing-ball modes.
In Sec.~\ref{sec:dyntun} we apply the fictitious integrable system approach 
\cite{BaeKetLoeSch2008,BaeKetLoe2010} to predict decay 
rates $\gamma_{mn}$, which describe the initial decay of bouncing-ball modes 
into the chaotic sea. Using the adiabatic modes as approximate bouncing-ball 
modes we evaluate Eq.~\eqref{eq:dyntun:fisa:gamma2} and find agreement to 
numerical rates for the cosine and the stadium billiard. 
For the cosine billiard, which has a corner of angle $\pi/2$ at $x=w$, we 
evaluate the fictitious integrable system approach analytically using the
eigenstates of the rectangular billiard as approximate bouncing-ball modes.
This leads to Eq.~\eqref{eq:dyntun:rect:gammares2} which shows excellent 
agreement with numerical rates. As a result we find that the decay rates 
of bouncing-ball modes of constant quantum number $n$ decrease as a power law
with increasing energy.

In Sec.~\ref{sec:countbb} we use the results on the decay rates in order 
to count the number of bouncing-ball modes $\Nbb(E)\sim E^\delta$ up to 
energy $E$. For the stadium billiard we recover the exponent $\delta = 3/4$
\cite{Tan1997,BaeSchSti1997}. For the cosine billiard we find 
$\delta\approx 0.64$. Using the analytical result
Eq.~\eqref{eq:dyntun:rect:gammares2} for the decay rates we derive 
$\delta = 5/8$, which is in agreement with our numerics and 
well below the previously predicted upper bound $\delta = 9/10$.


\begin{acknowledgments}
We are grateful to H.~Schanz, R.~Schubert, and G.~Tanner for stimulating 
discussions. Furthermore, we acknowledge financial support through the 
DFG Forschergruppe 760 ``Scattering systems with complex dynamics''.
\end{acknowledgments}



\begin{thebibliography}{37}

\bibitem{Sto1999}
 H.-J. St\"ockmann, \emph{Quantum Chaos. An introduction}
 (University Press, Cambridge, 1999).

\bibitem{NoeSto1997}
 J.~U. Noeckel and A.~D. Stone, Nature \textbf{385}, 45 (1997).

\bibitem{FriKapCarDav2001}
 N.~Friedman, A.~Kaplan, D.~Carasso, and N.~Davidson,
 Phys. Rev. Lett. \textbf{86}, 1518 (2001).

\bibitem{EllSchBer2001}
 C.~Ellegaard, L.~Schaadt, and P.~Bertelsen, Physica Skripta \textbf{T90},
 223 (2001).

\bibitem{MarRimWesHopGos1992}
 C.~M. Marcus, A.~J. Rimberg, R.~M. Westervelt, P.~F. Hopkins, and A.~C.
 Gossard, Phys. Rev. Lett. \textbf{69}, 506 (1992).

\bibitem{Bun1979}
 L.~A. Bunimovich, Commun. Math. Phys. \textbf{65}, 295 (1979).

\bibitem{Sin1970}
 Y.~G. Sinai, Russ. Math. Surveys \textbf{25}, 137 (1970).

\bibitem{Sti1996}
 P.~Stifter, Ph.D. thesis, Universit\"at Ulm (1996).

\bibitem{Hel1984}
 E.~J. Heller, Phys. Rev. Lett. \textbf{53}, 1515 (1984).

\bibitem{BaiHosSteTay1985}
 Y.~Y. Bai, G.~Hose, K.~Stefa\'nski, and H.~S. Taylor, Phys. Rev. A
 \textbf{31}, 2821 (1985).

\bibitem{OcoHel1988}
 P.~W. O'Connor and E.~J. Heller, Phys. Rev. Lett. \textbf{61},
 2288 (1988).

\bibitem{McdKau1988}
 S.~W. McDonald and A.~N. Kaufman, Phys. Rev. A \textbf{37}, 3067 (1988).

\bibitem{BogSch2004}
 E.~Bogomolny and C.~Schmit, Phys. Rev. Lett. \textbf{92}, 244102 (2004).

\bibitem{BurZwo2005}
 N.~Burq and M.~Zworski, SIAM Rev. \textbf{47}, 43 (2005).

\bibitem{Has2010}
 A.~Hassel, Ann. Math. \textbf{171}, 605 (2010).

\bibitem{Shn1974}
 A.~I. Shnirelman, Usp. Mat. Nauk \textbf{29}, 181 (1974).

\bibitem{Ver1985}
 Y.~Colin de Verdi\`ere, Commun. Math. Phys. \textbf{102}, 497 (1985).

\bibitem{Zel1987}
 S.~Zelditch, Duke. Math. J. \textbf{55}, 919 (1987).

\bibitem{GerLei1993}
 P.~G\'erard and E.~Leichtnam, Duke. Math. J. \textbf{71}, 559 (1987).

\bibitem{ZelZwo1996}
 S.~Zelditch and M.~Zworski, Comm. Math. Phys. \textbf{175}, 673 (1996).

\bibitem{Tan1997}
 G.~Tanner, J. Phys. A \textbf{30}, 2863 (1997).

\bibitem{BaeSchSti1997}
 A.~B\"acker, R.~Schubert, and P.~Stifter, J. Phys. A \textbf{30},
 6783 (1997).

\bibitem{BaeKetLoeSch2008}
 A.~B\"acker, R.~Ketzmerick, S.~L\"ock, and L.~Schilling,
 Phys. Rev. Lett. \textbf{100}, 104101 (2008).

\bibitem{BaeKetLoe2010}
 A.~B\"acker, R.~Ketzmerick, and S.~L\"ock, Phys. Rev. E \textbf{82},
 056208 (2010).

\bibitem{BaeKetLoeRobVidHoeKuhSto2008}
 A.~B\"acker, R.~Ketzmerick, S.~L\"ock, M.~Robnik, G.~Vidmar, R.~H\"ohmann,
 U.~Kuhl, and H.-J. St\"ockmann, Phys. Rev. Lett. \textbf{100},
 174103 (2008).

\bibitem{BaeKetLoeWieHen2009}
 A.~B\"acker, R.~Ketzmerick, S.~L\"ock, J.~Wiersig, and M.~Hentschel,
 Phys. Rev. A \textbf{79}, 063804 (2009).

\bibitem{LoeBaeKetSch2010}
 S.~L\"ock, A.~B\"acker, R.~Ketzmerick, and P.~Schlagheck,
 Phys. Rev. Lett. \textbf{104}, 114101 (2010).

\bibitem{DavHel1981}
 M.~J. Davis and E.~J. Heller, J. Chem. Phys. \textbf{75}, 246 (1981).

\bibitem{Ber1977}
 M.~V. Berry, J. Phys. A \textbf{10}, 2083 (1977).

\bibitem{BarBet2007}
 A.~H. Barnett and T.~Betcke, Chaos \textbf{17}, 043125 (2007).

\bibitem{HanOttAnt1984}
 J.~D. Hanson, E.~Ott, and T.~M. Antonsen, Phys. Rev. A \textbf{29}, 819 (1984).

\bibitem{PodNar2003}
 V.~A. Podolskiy and E.~E. Narimanov, Phys. Rev. Lett. \textbf{91}, 263601 
 (2003).

\bibitem{SheFisGuaReb2006}
 M.~Sheinman, S.~Fishman, I.~Guarneri, and L.~Rebuzzini,
 Phys. Rev. A \textbf{73}, 052110 (2006).

\bibitem{BetTre2005}
 T.~Betcke and L.~N. Trefethen, SIAM Rev. \textbf{47}, 469 (2005).

\bibitem{BaeKetLoeSch2011}
 A.~B\"acker, R.~Ketzmerick, S.~L\"ock, and H.~Schanz,
 Europhys. Lett. \textbf{94}, 30004 (2011).

\bibitem{BaeKetMon2005}
 A.~B\"acker, R.~Ketzmerick, and A.~G. Monastra, Phys. Rev. Lett.
 \textbf{94}, 054102 (2005).

\bibitem{BaeKetMon2007}
 A.~B\"acker, R.~Ketzmerick, and A.~G. Monastra, Phys. Rev. E
 \textbf{75}, 066204 (2007).

\end{thebibliography}
\end{document}